\documentclass[twocolumn,aps,amsmath,amssymb,pra]{revtex4-1}
\usepackage{bm}
\usepackage{xcolor}
\usepackage{graphicx}
\usepackage{boxedminipage}

\newcommand{\w}{\omega}
\newcommand{\wti}{\widetilde}

\newcommand{\B}{\mbox{\tiny B}}
\newcommand{\T}{\mbox{\tiny T}}
\newcommand{\tS}{\mbox{\tiny S}}
\newcommand{\s}{\mbox{\tiny S}}
\newcommand{\SB}{\mbox{\tiny SB}}
\newcommand{\AB}{\mbox{\tiny $AB$}}

\renewcommand{\d}{{\rm d}}

\newcommand{\dg}{\dagger}

\newcommand{\nl}{\nonumber \\}
\newcommand{\la}{\langle}
\newcommand{\ra}{\rangle}

\newcommand{\Sec}[1]{Sec.\,\ref{#1}}

\newcommand{\be}{\begin{equation}}
\newcommand{\ee}{\end{equation}}
\newcommand{\bea}{\begin{eqnarray}}
\newcommand{\eea}{\end{eqnarray}}
\newcommand{\bsube}{\begin{subequations}}
\newcommand{\esube}{\end{subequations}}
\newcommand{\Eq}[1]{Eq.\,(\ref{#1})}

\newcommand{\comments}[1]{}

\allowdisplaybreaks[1]

\begin{document}
\title{Nonequilibrium system--bath entanglement theorem versus heat transport}

\author{Peng-Li Du}\thanks{Authors of equal contributions}
\author{Zi-Hao Chen}\thanks{Authors of equal contributions}
\author{Yu Su}
\author{Yao Wang}%
\email{wy2010@ustc.edu.cn}
\author{Rui-Xue Xu} 
\author{YiJing Yan} \email{yanyj@ustc.edu.cn}

\affiliation{Hefei National Laboratory for Physical Sciences
 at the Microscale and Department of Chemical Physics
 and Synergetic Innovation Center of Quantum Information and Quantum Physics
 and Collaborative Innovation Center of Chemistry for Energy Materials
 ({\it i}{\rm ChEM}),
 University of Science and Technology of China, Hefei, Anhui 230026, China
}
	
\date{\today}

\begin{abstract}

 In this work, we extend the recently established
system--bath entanglement theorem (SBET) [J. Chem. Phys. {\bf 152}, 034102 (2020)] to
the nonequilibrium scenario,
in which an arbitrary system couples to
multiple Gaussian baths environments at different
temperatures.
While the existing SBET connects the entangled system--bath response
functions to those of local systems,
the extended theory is concerned with
the nonequilibrium steady--state 
quantum transport current
through molecular junctions.
The new theory is established on the basis of the generalized Langevin equation, with a close relation
to nonequilibrium thermodynamics
in the quantum regime.

\end{abstract}
\maketitle

\section{Introduction}

 Quantum transport of heat and particles
has attracted much attention in the past years.
On one hand, it is closely related to the fundamental
physics such as nonequilibrium thermodynamics in the quantum regime.
On the other hand, it also plays important roles 
in such as energy and quantum information applications.
Theoretical studies have been mainly carried
out in terms of nonequilibrium Green's function (NEGF) methods.\cite{Cho85118, Har08191}

 In this work, we exploit the well--established
system--bath entanglement theorem (SBET),\cite{Du20034102,Gon20214115}
with extension to nonequilibrium transport scenario.
Adopted here is the Gauss--Wick's environment ansatz\cite{Wei12,Yan05187}
that is commonly adopted in various quantum dissipation theories.
These include the formally exact
Feynman--Vernon influence functional theory,\cite{Fey63118}
and its derivative--equivalence the
hierarchical equations of motion (HEOM)
formalism.\cite{Tan906676,Tan06082001,Yan04216,Xu05041103,%
Xu07031107,Jin08234703,Zhe121129}
%
While the existing SBET
deals with for response functions only,\cite{Du20034102,Gon20214115}
the extended theory is concerned with the nonequilibrium steady--state
quantum transport current
through molecular junctions. In this context, the extended SBET provides an alternative approach to the NEGF formalism.
It is worth noting that the new theory is established on the basis of the generalized Langevin equation, which can readily support 
the evaluation on entangled system--bath correlation functions, 
which are closely related
to nonequilibrium thermodynamics
in the quantum regime.
The convention fluctuation--dissipation
theorem (FDT), which relates correlation functions
and response functions,
is only applicable to the equilibrium scenario.
There are no general relations between
the nonequilibrium correlation functions and
response functions.
It would be anticipated that the present Langevin equation based method be 
a viable approach toward such as fluctuation theorem far from equilibrium in the quantum regime.
For clarity, we focus on
the quantum heat transport formalism. The extension to electron current transport would be straightforward
 on the basis of the fermionic SBET.\cite{Gon20214115}

The remainder of this paper is organized as follows.
In \Sec{sec2}, we present the well--established SBET
for the response functions,\cite{Du20034102}
with extension
to the nonequilibrium transport scenario.
In \Sec{sec3}, we construct a novel SBET, on the basis of a generalized Langevin equation, which readily leads to NEGF formalism for the quantum heat transport current.
We conclude this work to the end of \Sec{sec3}.

\section{Extended system--bath entanglement theorem}
\label{sec2}

\subsection{Langevin equation for hybrid bath dynamics}
\label{thsec2A}

 System--bath entanglement plays a crucial role
in dynamic and thermal properties of complex systems.
This is concerned with a currently active topic in quantum mechanics of open systems. Recently, we had constructed the SBET.\cite{Du20034102,Gon20214115}
This theorem comprises exact relations between
the entangled system--bath response functions and those of local
anharmonic systems.
Applications had been demonstrated 
with Fano interference spectroscopy.\cite{Du20034102}
The SBET had also been exploited in the establishment of
the thermodynamic free--energy
spectrum theory.\cite{Gon20214115}

To extend this theory to the nonequilibrium scenario,
we should include multiple bath reservoirs with different temperatures,
so that heat transport is anticipated.
The total system--and--bath composite Hamiltonian reads
\be \label{Hall}
 H_{\T}=H_{\tS}+h_{\B} + H_{\tS\B}
=H_{\tS} +\sum_{\alpha}h_{\alpha}+\sum_{\alpha u}\hat Q_{u}\hat F_{\alpha u}.
\ee
The system Hamiltonian $H_{\s}$ and dissipative
modes $\{\hat Q_u\}$ are arbitrary.
The $\alpha$-reservoir bath Hamiltonian
and the hybrid bath modes are modelled with
\be\label{gwmodel}
 h_{\alpha}
=\frac{1}{2}\sum_j\omega_{\alpha j}(\hat p_{\alpha j}^2+\hat x_{\alpha j}^2)
\ \ {\rm and}\  \
 \hat F_{\alpha u}=\sum_j c_{\alpha u j}\hat x_{\alpha j},
\ee
respectively, which together constitute
the so--called Gauss--Wick's environment.\cite{Wei12,Yan05187}
The simplicity arises from the fact that
the interacting bath commutators are all c--variables;
i.e.,
\be\label{cnumber}
\phi_{uv}^{\alpha}(t)
\equiv 
i[\hat F^{\B}_{u}(t),\hat F^{\B}_v(0)]
=\sum_j c_{\alpha u j}c_{\alpha v j}\sin(\w_{\alpha i}t)
\ee
where $\hat F^{\B}_{\alpha u}(t)\equiv e^{ih_{\B}t}\hat F_{\alpha u} e^{-ih_{\B}t}=
e^{ih_{\alpha}t}\hat F_{\alpha u} e^{-ih_{\alpha}t}$.
Throughout the paper we set
$\hbar=1$ and
$\beta_{\alpha}=1/(k_BT_{\alpha})$, with $k_B$ being
the Boltzmann constant and $T_{\alpha}$ the $\alpha$--reservoir temperature.

Denote also $\hat O(t)\equiv e^{iH_{\T}t}\hat Oe^{-iH_{\T}t}$,
with noticing that $\hat F_{\alpha u}(t)\neq \hat F^{\B}_{\alpha u}(t)$.
The former is defined via the total system--and--bath composite space,
whereas the latter is a bare bath subspace property. 
It is easy to obtain \cite{Du20034102}
\be\label{eq07}
   \hat F_{\alpha u}(t)=\hat F^{\B}_{\alpha u}(t)-\sum_v\int_{t_0}^t\!\!{\rm d}\tau\,\phi_{uv}^{\alpha}(t-\tau)\hat Q_v(\tau).
\ee
Note that $\phi_{uv}^{\alpha}(t)$, \Eq{cnumber}, 
can be recast as
\be\label{phiB_t}
 \phi_{uv}^{\alpha}(t)=i\la 
 [\hat F^{\B}_{\alpha u}(t),\hat F^{\B}_{\alpha v}(0)]\ra_{\alpha},
\ee
with $\la(\,\cdot\,)\ra_{\alpha}\equiv 
{\rm tr}_{\B}[(\,\cdot\,)e^{-\beta_{\alpha}h_{\alpha}}]
/{\rm tr}_{\B}e^{-\beta_{\alpha}h_{\alpha}}$.
The hybridization bath spectral density is given by\cite{Zhe121129,Yan16110306}
\be\label{Jw}
 J^{\alpha}_{uv}(\w) \equiv 
\frac{1}{2}\int^{\infty}_{-\infty}\!\!\d t\,
 e^{i\w t}\la [\hat F^{\B}_{\alpha u}(t),\hat F^{\B}_{\alpha v}(0)]\ra_{\alpha}.
\ee
Its microscopic equivalence 
reads [cf.\,\Eq{gwmodel}]
\be\label{Jw0}
 J^{\alpha}_{uv}(\w)=\frac{\pi}{2}\sum_{j}
  c_{\alpha uj}c_{\alpha vj}
  [\delta(\w-\w_{\alpha j})-\delta(\w+\w_{\alpha j})].
\ee 
Evidently, $J^{\alpha}_{uv}(\w)=J^{\alpha}_{vu}(\w)
=- J^{\alpha}_{uv}(-\w)$.

  It is worth noting that 
the Langevin equation (\ref{eq07}), together 
with the property of \Eq{cnumber},
will give rise to some interesting relations
between the entangled system--bath properties
and the local system ones, as bridged with
the bare--bath $\phi^{\alpha}_{uv}(t)$ or
$J^{\alpha}_{uv}(\w)$.

\subsection{The system--bath entanglement theorem
 for response functions and expectation values}
 
 The SBET is a type of input--output formalism,
in which the local system properties, such as
\be
 \chi^{\s\s}_{uv}(t)\equiv i\la[\hat{Q}_u(t),\hat{Q}_v(0)]\ra 
\ee 
are the input functions, whereas
the nonlocal correspondences,
\begin{align}\label{eq11}
\begin{split}
  \chi^{\s\alpha}_{uv}(t)&\equiv i\la[\hat{Q}_u(t),\hat{F}_{\alpha v}(0)]\ra
\\
  \chi^{\alpha\s}_{u v}(t)&\equiv i\la[\hat{F}_{\alpha u}(t),\hat{Q}_{v}(0)]\ra
\end{split}
\intertext{and}   
 \chi^{\alpha\alpha'}_{uv}(t)
 &\equiv i\la[\hat{F}_{\alpha u}(t),\hat{F}_{\alpha' v}(0)]\ra
\end{align}
are the output functions.
Here, 
\be
 \chi_{\AB}(t-\tau)\equiv i\la[\hat A(t),\hat B(\tau)]\ra
\ee
are defined in the total composite space, 
at nonequilibrium steady--state scenario, and $\la \,\cdot \,\ra$ denotes the ensemble average over the total composite space steady--state density operator.
It is easily to verify that the established
SBET does include the 
general nonequilibrium scenario.\cite{Du20034102} 
The final results, in terms of
the matrices, are  
\be\label{eq12}
\begin{split}
  {\bm\chi}^{\alpha\tS}(t)
&=-\int_0^t\!{\rm d}\tau\,
   {\bm\phi}^{\alpha}(t-\tau){\bm\chi}^{\tS\tS}(\tau),
\\
  {\bm\chi}^{\tS\alpha}(t)
&=-\int_0^t\!{\rm d}\tau\,
   {\bm\chi}^{\tS\tS}(\tau){\bm\phi}^{\alpha}(t-\tau),
\end{split}
\ee
and
\begin{align}\label{eq13}
  {\bm\chi}^{\alpha\alpha'}(t)
&=\int_0^t\!{\rm d}\tau\!\int^{\tau}_{0}\!\d\tau'\,
   {\bm\phi}^{\alpha}(t-\tau){\bm\chi}^{\tS\tS}(\tau')
   {\bm\phi}^{\alpha'}(\tau-\tau')
\nl&\quad
  +\delta_{\alpha\alpha'}{\bm\phi}^{\alpha}(t).
\end{align}
In the frequency domain, 
$\wti f(\omega)=\int_0^\infty {\rm d}t\,e^{i\omega t}f(t)$, 
the above expressions read 
\be\label{eq15}
\begin{split}
&\wti{\bm \chi}^{\alpha \tS}(\w)=-\wti{\bm \phi}^{\alpha}(\w)\wti{\bm \chi}^{\tS\tS}(\w),
\\
&\wti{\bm \chi}^{\tS\alpha }(\w)=-\wti{\bm \chi}^{\tS\tS}(\w)\wti{\bm \phi}^{\alpha}(\w),
\end{split}
\ee
and
\be
 \wti{\bm\chi}^{\alpha\alpha'}(\w)
=\wti{\bm\phi}^{\alpha}(\w)\wti{\bm\chi}^{\tS\tS}(\w)
   \wti{\bm\phi}^{\alpha'}(\w)
  +\delta_{\alpha\alpha'}\wti{\bm\phi}^{\alpha}(\w).
\ee
Moreover, \Eq{eq07} will also give rise to the expectation
values the following input--output relations,\cite{Gon20214115}
\be\label{ggg}
 \la \hat F_{\alpha u}\ra=-\sum_{v}\eta^{\alpha}_{uv}\la \hat Q_v\ra,
\ee
where
\be\label{eta_def}
 \eta_{uv}^{\alpha}\equiv \int_{0}^{\infty}\!\!{\rm d}t\,\phi_{uv}^{\alpha}(t).
\ee

\section{Onset of heat current}
\label{sec3}

\subsection{Heat current}

 Let us start with the heat current
transferring from the specified $\alpha$--reservoir
to the central system. The related current
operator would read [cf.\,\Eq{Hall} with \Eq{gwmodel}]
\begin{align}\label{currop}
  \hat{J}_{\alpha} \equiv -\frac{{\rm d}h_{\alpha}}{{\rm d}t}
=-i[H_{\T},h_{\alpha}]
=\sum_u\hat Q_u \dot{\hat F}_{\alpha u}.
\end{align}
It is noticed there is another convention of heat current operator definition that engages  the hybrid bath modes of $\hat F_{\alpha u}$ only.
\cite{Son17064308,Esp15235440,Sch15224303}
Others are just linear combinations of above two definitions. The existing dissipaton equation of motion theory can be exploited to the direct evaluation on the transport current and the noise spectrum.\cite{Yan14054105,Zha18780,Wan20041102}

 The quantity of interest in this section is
\be\label{Jave}
 J_{\alpha} \equiv \la \hat J_{\alpha}\ra
=\sum_u \la\hat Q_u \dot{\hat F}_{\alpha u}\ra.
\ee
The direct evaluation can be carried out by exploiting
the established dissipaton equation of motion (DEOM) theory.\cite{Wan20041102}
 In the following, we will establish
the extended SBET for the indirect evaluation
of \Eq{Jave}. The new theory can be numerically
validated with respect to the aforementioned direct
evaluations; See \Sec{thsec3c}.

\subsection{The extended system--bath entanglement theory}

 It is noticed that the transport current
consists of absorptive ($\w>0$) and emissive ($\w<0$) components.
In this contact, we decompose the 
hybrid bath operator, $\hat F_{\alpha u}$ in \Eq{gwmodel} as
\be
 \hat F_{\alpha u}=\sum_{\sigma=+,-}\hat F_{\alpha u}^{\sigma}.
\ee
Mathematically,  $\hat F_{\alpha u}^{\pm}$
comprises the linear combinations
of the  creation/annihilation operators 
associated with the effective bath modes in the canonical
ensembles.\cite{Ume95}
In parallel, \Eq{eq07} is decomposed into its components,
\be \label{hhh}
 \hat F_{\alpha u}^{\sigma}(t)
=\hat F_{\alpha u}^{\B;\sigma}(t)-\sum_v\int_{t_0}^t\!\!{\rm d}\tau\,\phi_{uv}^{\alpha;\sigma}(t-\tau)\hat Q_v(\tau).
\ee
The involving $\phi_{uv}^{\alpha;\sigma}(t)$ satisfies not only
\bsube\label{eq22}
\be
 \phi_{uv}^{\alpha;+}(t)+\phi_{uv}^{\alpha;-}(t)
=\phi_{uv}^{\alpha}(t),
\ee
but also
\begin{align}
&\quad\,\phi_{uv}^{\alpha;+}(t)-\phi_{uv}^{\alpha;-}(t)
\nl
&=\frac{2}{i\pi}\!\int^{\infty}_{0}\!\!\d\w \cos(\w t)
 \coth(\beta_{\alpha}\w/2)J_{uv}(\w),
\end{align}
\esube
for the required canonical ensemble properties.

To compute the heat current, \Eq{Jave}, 
with \Eq{hhh}, we have
\begin{align}\label{ggg}
 \dot{\hat F}^{\sigma}_{\alpha u}(t)&=\dot{\hat F}^{\B;\sigma}_{\alpha u}(t)-\sum_v\int_{t_0}^t\!\!{\rm d}\tau\,\dot\phi_{uv}^{\alpha;\sigma}(t-\tau)\hat Q_v(\tau)
 \nl &\quad
 -\sum_{v}\phi_{uv}^{\alpha;\sigma}(0) \hat Q_v(t).
\end{align}
Moreover, the identities ${\hat F}^{+}_{\alpha u}
=({\hat F}^{-}_{\alpha u})^{\dg}$ and
$[{\hat F}^{\sigma}_{\alpha u}, \hat Q_v]=0$
result in
\be \label{sss}
\la \hat Q_u \dot{\hat F}_{\alpha u}\ra
=\sum_{\sigma=+,-}\la \dot{\hat F}^{\sigma}_{\alpha u}\hat Q_{u} \ra
=\la \dot{\hat F}^{+}_{\alpha u}\hat Q_u \ra+{\rm c.c.}
\ee
Now, it is readily to obtain
\be\label{eq26}
\la \hat Q_u \dot{\hat F}_{\alpha u}\ra
=-2{\rm Re}\sum_{v}\!\int_{0}^{\infty}\!\!{\rm d}\tau\,
 \dot\phi_{uv}^{\alpha;+}(\tau)\la\hat Q_{v}(0)\hat Q_u(\tau)\ra.
\ee
The involving $\dot\phi_{uv}^{\alpha;+}(\tau)$ 
is determined via \Eq{eq22}.
Simple algebra then gives rise to the transport current the final result,
\begin{align}\label{gfr}
J_{\alpha}&
=\frac{2}{\pi}\sum_{uv}\!\int_{-\infty}^{\infty}\!\!\!{\rm d}\w\,\frac{\w}{e^{\beta_{\alpha}\w}-1} J^{\alpha}_{uv}(\w)C_{vu}(\w)
,
\end{align}
where 
\be\label{spec}
C_{vu}(\w)\equiv\frac{1}{2}\int_{-\infty}^{\infty}\!\!{\d}t\,e^{i\w t} \la\hat Q_{v}(t)\hat Q_u(0)\ra.
\ee
It is easy to show that \Eq{gfr} is identical to the Meir--Wingreen's NEGF formalism.\cite{Mei922512}

\subsection{Numerical validations and concluding remarks}\label{thsec3c}

For illustrations, consider the total composite Hamiltonian, $H_{\T}$ of \Eq{Hall}, with 
\be 
H_{\tS}=V(|1\ra\la 2|+|2\ra\la 1|),
\ee
$h_{\B}=h_{\rm L}+h_{\rm R}$ and 
\be 
H_{\SB}=\sum_{u=1,2}|u\ra\la u|(\hat F_{{\rm L }u}+\hat F_{{\rm R }u}).
\ee
Evidently, $\hat Q_u=|u\ra\la u|$.
%
Adopt further
\be
\wti\phi^{\alpha}_{uv}(\w)=\delta_{uv}\frac{\eta^{\alpha}_{u}\Omega^2}{\Omega^2-\w^2-i\w\zeta}.
\ee
Set $\eta^{\rm L}_{1}=\eta^{\rm R}_{2}=0.2V$, $\eta^{\rm R}_{1}=\eta^{\rm L}_{2}=0.4V$, $\Omega=2V$, $\zeta=10V$ and $k_{B}T_{\rm L}=5V$.
Table \ref{table1} reports the results of numerical validation at the specified 
values of $T_{\rm R}/T_{\rm L}$. As mentioned after \Eq{Jave}, the direct evaluation refers to the DEOM results, whereas the indirect ones arise from \Eq{gfr}, through the local system spectra, \Eq{spec}.  The extended SBET, \Eq{gfr}, does hold for arbitary systems in the nonequilibrium steady--state scenario.

\begin{table}[htbp!] 	
	\begin{tabular}{|c|c|c|c|c|}
		\hline
		$T_{\rm R}/T_{\rm L}$&0.5&\ \ 1\ \ \ &1.5&2 \\
		\hline
		Direct & 0.01484 & \ \ 0\ \ \ & $-0.008757$ & $-0.01435$ \\
		\hline
		Indirect &  0.01487& \ \ 0\ \ \  & $-0.008773$& $-0.01435$ \\
		\hline
	\end{tabular}
	\caption{Direct versus indirect approach to the heat current $J_{\rm L}$, as expressed in \Eq{gfr}.}\label{table1}
\end{table}

In summary, we revisit the NEGF formalism via the generalized Langevin equation (\ref{eq07}). The present approach can be readily extended to the entangled system--bath correlation functions that would be closely related to  nonequilibrium thermodynamics in the quantum regime.

\begin{acknowledgments}
  Support from
  the Ministry of Science and Technology No.\ 2017YFA0204904
  and the Natural Science Foundation of China No.\ 21633006
  is gratefully acknowledged.
\end{acknowledgments}


\end{document}